\documentclass[12pt]{book} 
\usepackage[sectionbib]{natbib}
\usepackage{array,epsfig,fancyheadings,rotating}
\usepackage[]{hyperref}  

\usepackage{amsmath}
\usepackage{amssymb}
\usepackage{amsfonts}
\usepackage{multirow}
\usepackage{amsthm}

\theoremstyle{definition}

\newcommand{\E}{\mathrm{E}}
\newcommand{\tr}{\mathrm{tr}}
\newcommand{\diag}{\mathrm{diag}}
\usepackage[plain,noend,boxed]{algorithm2e}

\def\thesection{\arabic{section}}
\begin{document}


\renewcommand{\baselinestretch}{2}


\markboth{\hfill{\footnotesize\rm LIUYI HU, WENBIN LU, JIN ZHOU AND HUA ZHOU} \hfill}
{\hfill {\footnotesize\rm MM ALGORITHMS FOR LOGISTIC LINEAR MIXED MODEL} \hfill}

\renewcommand{\thefootnote}{}
$\ $\par


\fontsize{12}{14pt plus.8pt minus .6pt}\selectfont \vspace{0.8pc}
\centerline{\large\bf MM ALGORITHMS FOR VARIANCE COMPONENT ESTIMATION}
\vspace{2pt} \centerline{\large\bf AND SELECTION IN LOGISTIC LINEAR MIXED MODEL}
\vspace{.4cm} \centerline{Liuyi Hu$^1$, Wenbin Lu$^1$, Jin Zhou$^2$ and Hua Zhou$^3$} \vspace{.4cm} \centerline{\it
$^1$North Carolina State University, $^2$University of Arizona, and $^3$University of California, 
Los Angeles} \vspace{.55cm} \fontsize{9}{11.5pt plus.8pt minus
.6pt}\selectfont


\begin{quotation}
\noindent {\it Abstract:}
Logistic linear mixed model is widely used in experimental designs and genetic analysis with binary traits. Motivated by modern applications, we consider the case with many groups of random effects and each group corresponds to a variance component. When the number of variance components is large, fitting the logistic linear mixed model is challenging. We develop two efficient and stable minorization-maximization (MM) algorithms for the estimation of variance components based on the Laplace approximation of the logistic model. One of them leads to a simple iterative soft-thresholding algorithm for variance component selection using maximum penalized approximated likelihood. We demonstrate the variance component estimation and selection performance of our algorithms by simulation studies and a real data analysis.\par

\vspace{9pt}
\par
\end{quotation}\par

\def\thefigure{\arabic{figure}}
\def\thetable{\arabic{table}}

\renewcommand{\theequation}{\thesection.\arabic{equation}}

\fontsize{12}{14pt plus.8pt minus .6pt}\selectfont

\setcounter{section}{1} 
\setcounter{equation}{0} 
\noindent {\bf 1. Introduction}\label{sec:intro}


Generalized linear mixed model (GLMM) is an extension of generalized linear model to incorporate random effects accounting for heterogeneity among responses \citep{Mcculloch2001GLMM, Stroup2012GLMM}. It is widely used in clustered, longitudinal, and panel data analysis \citep{Zeger1991GLMM, BreslowClayton93GLMM}. Logistic linear mixed model is one of the GLMMs for binary responses and assumes
\begin{equation}\label{eqn:lvc-model}
\begin{aligned}
y_j \mid \eta_j &\sim \text{Bernoulli}(\mu_j) \\
\mu_j &= 1 / \left\{1 + \exp(-\eta_j)\right\} \\
\end{aligned}
\end{equation}
for $j = 1, \dots, n$ and $\eta = (\eta_1, \ldots, \eta_n)^T$ takes the form 
$$
\eta = X \beta + Z_1 u_1 + \cdots + Z_m u_m,
$$ 
where $X$ and $Z = \left(Z_1, \dots, Z_m\right)$ are known predictor matrices, $\beta$ is the coefficient vector for fixed effects, and $u_i \sim N(0_{q_i}, \sigma_i^2 I_{q_i})$ are independent random effects. Because
\begin{eqnarray*}
	\eta	\sim N(X \beta, \sigma_1^2 Z_1 Z_1^T + \cdots + \sigma_m^2 Z_m Z_m^T),
\end{eqnarray*}
we call $\sigma_1^2, \ldots, \sigma_m^2$ variance components. 

Logistic linear mixed model finds applications in agriculture, econometrics, biology and genetics. Two motivating examples are the analysis of variance (ANOVA) for dichotomous responses \citep{Anderson1985VC, Quene2008ANOVA} and the quantitative trait loci (QTL) mapping for binary traits \citep{Yi1999QTLmapping, Che2012QTLGLMM}. In ANOVA, $Z_i$ corresponds to each factor or their interactions. In modern applications, the number of factors can be large and the number of interaction terms increases quadratically with the number of factors. In QTL mapping, $Z_i$ corresponds to a gene region. The number of genes $m$ is at order of $10^2 \sim 10^3$ in a typical genetic study.
In Section 4 and 5, we will discuss further about these two applications as well as associated analysis using our proposed algorithms. 

In general direct maximization of the GLMM likelihood function is computationally intractable because it involves potentially high-dimensional integrals. The existing methods involve various forms of approximations. The first class of methods use numerical integration such as Gaussian quadrature \citep{Davidian1992GaussianQuadrature} and adaptive Gaussian quadrature \citep{PinheiroBates1995ApproximationsNLM}. These methods are applicable only to low dimensional integrals and thus limited to problems where data form very small independent clusters. The second type of methods invoke the Laplace approximation \citep{,Wolfinger93LA,ShunMcCullagh95LA} or its variants such as the penalized quasi-likelihood \citep{BreslowClayton93GLMM} and the integrated nested Laplace approximation \citep{RueMartinoChopin09INLA}. The third class of methods resort to Monte Carlo methods to approximate either the original integral \citep{SungGeyer07MCMLE} or the E step of EM algorithm \citep{BoothHobert99GLMM-MCEM}. \cite{PinheiroBates1995ApproximationsNLM} compare and discuss penalized quasi-likelihood (PQL), Laplace approximation, importance sampling, Gaussian quadrature, and adaptive Gaussian quadrature (AGQ). They conclude that Laplace approximation and adaptive Gaussian quadrature give the ``best mix of efficiency and accuracy". In this paper, we propose algorithms based on the Laplace approximation of the log-likelihood function because AGQ is numerically infeasible for the ANOVA and genetic applications we are considering.

Our primary interest is in the estimation and selection of variance components. Researchers have worked on selecting fixed effects in GLMMs \citep{Groll2014VSGLMM, Schelldorfer2014GLMMlasso}. For random effects selection, however, most procedures are developed in the framework of linear mixed models \citep{Bondell2010JointLMM, AhnZhangLu012MomentRFLMM} for quantitative responses. In contrast only few references discuss random effects selection in GLMM. \cite{Ibrahim2011SelectMixedEffect} develop a simultaneous fixed and random effects selection procedure based on the SCAD and adaptive LASSO penalties using a Monte Carlo EM for general mixed models.  \cite{CaiDunson2006BayesianCovSelection} propose a method for random effect selection in GLMMs within the Bayesian framework using a stochastic search MCMC algorithm. \cite{Pan2014RandomGLMM} propose a backfitting algorithm  to select effective random effects based on penalized quasi-likelihood (PQL) function. However all the above mentioned papers study the clustered data with repeated measurements on the subjects. They assume $n$ independent subjects with observations $\left(y_1, X_1, Z_1 \right), \dots, \left(y_n, X_n, Z_n \right)$ and 
\begin{equation}\label{eqn:cluster-model}
\E\left(y_i \mid X_i, Z_i, b_i\right) = g\left(\eta_i\right) = g\left(X_i\beta + Z_ib_i\right),
\end{equation}
where $g(\cdot)$ is some known link function, $X_i$ and $Z_i$ are known matrices and $b_i \sim N_q\left(0, D\right)$ is the random effect. 
Here, $D$ is the unknown covariance matrix shared by the subjects that is to be estimated by maximizing some penalized likelihood. For example,  \cite{Ibrahim2011SelectMixedEffect} perform the penalization on the Cholesky decomposition of $D$, denoted as $\Gamma$, such that each row of $\Gamma$ either are all not zero  or all zero and \cite{Pan2014RandomGLMM} penalize on positive elements proportional to the standard deviation of the random effects $b_i$.   
In this paper, we propose an algorithm for selection of random effects by shrinking the variances of ineffective random effects towards zero based on penalized likelihood defined in Section 3.3. There are two key differences between our variance components selection and previous work.  First,  model \eqref{eqn:cluster-model} is not the same as model \eqref{eqn:lvc-model} we want to address in this paper. Model \eqref{eqn:lvc-model} can deal with clustered data (like ANOVA) but not restricted to it and it assumes that the random effects $u_i \sim N(0_{q_i}, \sigma_i^2 I_{q_i})$ are independent. Second, the random effects selection on model \eqref{eqn:cluster-model} is selecting individual random effect while for model \eqref{eqn:lvc-model} we are selecting groups of random effects, i.e. the random effects in each $u_i$ are either all selected or not. To the best of our knowledge, there exist no literature about variance components selection for model \eqref{eqn:lvc-model}. 


In this paper, based on the minorization-maximization (MM) principle \citep{Lange00OptTrans}, we propose two novel algorithms for variance component estimation under two different parameterizations of logistic linear mixed model and then extend to variance component selection by incorporating penalization. The first parameterization is efficient for estimating parameters without penalty, while the second easily generalizes to penalized estimation. Both algorithms are simple to implement and numerically stable. Our simulation studies and real data analysis demonstrate that the proposed algorithms outperform the commonly used tools and are scalable to high-dimensional problems.

\setcounter{section}{2} 
\setcounter{equation}{0} 
\noindent {\bf 2. Preliminaries}

Throughout we reserve Greek letters for parameters and indicate the current iteration number by a superscript $t$. 

\noindent {\bf 2.1. The MM principle}

The MM principle \citep{Lange00OptTrans, Hunter2004MMtutorial} for maximizing an objective function $f(\theta)$ involves two M-steps. The first M-step minorizes the objective function $f(\theta)$ by a surrogate function $g(\theta \mid \theta^{(t)})$ at the current iterate $\theta^{(t)}$.  Minorization is a combination of a tangent condition $f(\theta^{(t)})  =  g(\theta^{(t)} \mid \theta^{(t)})$ and a domination condition $f(\theta)  \ge  g(\theta \mid \theta^{(t)})$ for $ \theta \ne \theta^{(t)}$. 
The second M-step is defined by the iterates:
	\begin{equation}
		\theta^{(t+1)} = \underset{\theta} {\text{arg max}} \ g(\theta \mid 			\theta^{(t)}).
 	\end{equation}
Because 
	\begin{equation}
		\begin{aligned}
		f(\theta^{(t+1)})  \ge  g(\theta^{(t+1)} \mid \theta^{(t)}) \ge 	 g(\theta^{(t)} \mid \theta^{(t)}) =  f(\theta^{(t)}), 					\label{eqn:MM-monotonicity}
		\end{aligned}
	\end{equation} 
the MM iterates satisfy the
ascent property, which drives the objective function uphill and makes the MM algorithm remarkably stable. 

Our derivation of the MM algorithms for variance components estimation and selection hinges on two minorizations.

\noindent {\bf 2.2. Supporting hyperplane minorization}

If $f(\theta)$ is convex and differentiable, then the supporting hyperplane
\begin{equation}\label{eqn:support-hyper}
g(\theta)  =   f(\theta^{(t)}) + \nabla f(\theta^{(t)})^T(\theta - \theta^{(t)})
\end{equation}
is a minorization function of $f(\theta)$ at $\theta^{(t)}$ \citep{Hunter2004MMtutorial}.

For symmetric matrices we write $A \preceq B$ when $B - A$ is positive semidefinite. A matrix-valued function $f$ is said to be (matrix) convex if 
\begin{eqnarray*}
	f\left\{\lambda A + (1-\lambda) B\right\} & \preceq & \lambda f(A) + (1-\lambda) f(B)
\end{eqnarray*}
for all $A$, $B$, and $\lambda \in [0,1]$.  Since the negative log determinant function $f(B) = -\log \det B$ is convex on the set of positive definite matrices \citep{BoydVandenberghe04Book} and the supporting hyperplane of $f(B)$  is 
\begin{eqnarray*}
g(B) &=& f(B^{(t)}) +  \nabla f(B^{(t)})^T(B - B^{(t)}) \\
&=&-\log \det B^{(t)} - \tr \left\{\left(B^{(t)}\right)^{-1}\left(B - B^{(t)}\right)\right\},
\end{eqnarray*}
the supporting hyperplane minorization described above yields the following inequality
\begin{equation}\label{eqn:logdet-minorization}
	-\log \det B  \geq  -\log \det B^{(t)} - \tr \left\{\left(B^{(t)}\right)^{-1}\left(B - B^{(t)}\right)\right\}.
\end{equation}  

\noindent {\bf 2.3. Quadratic minorization}

If a convex function $f(\theta)$ is twice differentiable and there exists a matrix $M$ such that $M \preceq \nabla^2 f(\theta)$ for all $\theta$, then 
\begin{equation}\label{eqn:quad-minorization}
g(\theta)   =   f(\theta^{(t)}) + \nabla f(\theta^{(t)})^T(\theta - \theta^{(t)}) + \frac 12 (\theta - \theta^{(t)})^{T}M (\theta - \theta^{(t)})
\end{equation}
is a minorization function of $f(\theta)$ at $\theta^{(t)}$ \citep{Hunter2004MMtutorial}.


\lhead[\footnotesize\thepage\fancyplain{}\leftmark]{}\rhead[]{\fancyplain{}\rightmark\footnotesize\thepage}

\setcounter{section}{3} 
\setcounter{equation}{0} 
\noindent {\bf 3. Algorithms for estimation}

\noindent {\bf 3.1. Model formulation 1}

The likelihood  for model \eqref{eqn:lvc-model} is 
	\begin{equation} \label{eqn:likelihood}
		L(\beta, \sigma) = \int \exp \{h(u \mid \beta, \sigma^2) \} \, du,
	\end{equation}
where $\sigma = \left(\sigma_1, \dots, \sigma_m\right)^T$ with $\sigma_i \geq 0 $ for $i=1, \dots, m$, $\sigma^2 = \left(\sigma_1^2, \dots, \sigma^2_m\right)^T$ and the complete log-likelihood is 
	\begin{eqnarray*}
		h(u \mid \beta, \sigma^2)
		&=& \sum_j \left\{y_j \eta_j - \ln (1 + e^{\eta_j})\right\} - \frac 12 \sum_{i=1}^m \left( q_i \ln \sigma_i^2 + \frac{\|u_i\|_2^2}{\sigma_i^2} \right) \\
		&=& \sum_j \left\{y_j \eta_j - \ln (1 + e^{\eta_j})\right\} - \frac 12 \sum_{i=1}^m \frac{\|u_i\|_2^2}{\sigma_i^2} + \text{ terms without } u_i.
	\end{eqnarray*}
	
Direct optimization of the likelihood defined in \eqref{eqn:likelihood} is computationally challenging because of the integral. The Laplace approximation (LA) to the likelihood $L(\beta, \sigma)$ is obtained by replacing $h(u \mid \beta, \sigma^2)$ by its second-order Taylor expansion at the conditional maximum. Given current iterate $(\beta, \sigma)$, let $u^*$ be the maximizer of $h$ and $\eta^* = X \beta + Z u^*$ where $Z = \left(Z_1, Z_2, \dots, Z_m\right)$. Then the approximated log-likelihood is
	\begin{eqnarray}
		L_{\text{LA}}(\beta, \sigma)
		&=& h(u^* \mid \beta, \sigma^2) - \frac 12 \ln \det \nabla^2 \left\{- h(u^* \mid \beta, \sigma^{2})\right\} \nonumber\\
		&=& \sum_j \left\{y_j \eta_j^* - \ln \left(1 + e^{\eta_j^*} \right) \right\} - \frac 12 \sum_{i=1}^m q_i \ln \sigma_i^2 - \frac 12 \sum_{i=1}^m \frac{\|u_i^*\|_2^2}{\sigma_i^2} \nonumber\\
		& & \quad - \frac 12 \ln \det \left\{ Z^T W^* Z + \text{blkdiag}(\sigma_1^{-2}I_{q_1}, \ldots, \sigma_m^{-2} I_{q_m}) \right\} \nonumber\\
		&=& \sum_j \left\{y_j \eta_j^* - \ln \left(1 + e^{\eta_j^*} \right) \right\}- \frac 12 \sum_{i=1}^m \frac{\|u_i^*\|_2^2}{\sigma_i^2} \label{eqn:laplace-likelihood}\\
		& & \quad  - \frac 12 \ln \det \left( W^{*-1} + \sum_i \sigma_i^2 Z_i Z_i^T \right) - \frac 12 \ln \det W^* \nonumber\\
		& & \quad + \text{ terms without } \beta , \sigma^2,
\nonumber	\end{eqnarray}
where $W^* = \diag(w^*)$ is a diagonal matrix with entries
	\begin{eqnarray*}
		w^*_j = p^*_j (1 - p^*_j) = \frac{e^{\eta_j^*}}{\left( 1 + e^{\eta_j^*} \right)^2} \ \text{ and } \ p^*_j = \frac{e^{\eta_j^*}}{\left(1+ e^{\eta_j^*}\right)}. 
	\end{eqnarray*}
Detailed derivations of the approximated log-likelihood \eqref{eqn:laplace-likelihood} are in the Appendix.
The MM algorithm cycles through following updates of $u$, $\beta$ and $\sigma^2$.
\begin{enumerate}

\item To maximize $h(u \mid \beta, \sigma^2)$, the gradient and Hessian are
	\begin{eqnarray*}
		\nabla_{u} h &=& Z^T (y - p) - \begin{pmatrix} \sigma_1^{-2} u_1 \\ \vdots \\ \sigma_m^{-2} u_m \end{pmatrix} \\
		\nabla_{u}^2 h &=& - \left\{ Z^T WZ + \text{blkdiag}(\sigma_1^{-2}I_{q_1}, \ldots, \sigma_m^{-2} I_{q_m}) \right\},
	\end{eqnarray*}
where $p = \left(p_1, \dots, p_n\right)^T$ with $p_j = e^{\eta_j} / (1 + e^{\eta_j})$ and $W = \diag (w_1, \dots, w_n)$ with $w_j = p_j(1-p_j)$. 
Since each $w_j$ is upper bounded by 0.25, it follows that 
$$
\nabla_{u}^2 h  \succeq  - \left\{ 0.25 Z^TZ + \text{blkdiag}(\sigma_1^{-2}I_{q_1}, \ldots, \sigma_m^{-2} I_{q_m}) \right\}.
$$
Thus we can construct a quadratic minorization function at $u^{(l)}$ using \eqref{eqn:quad-minorization} and maximizing the quadratic surrogate gives the MM update
	\begin{eqnarray}\label{eqn:model1-maximize-mu}
		u^{(l+1)}  =  u^{(l)} + \left\{ 0.25 Z^TZ + \text{blkdiag}(\sigma_1^{-2}I_{q_1}, \ldots, \sigma_m^{-2} I_{q_m}) \right\}^{-1} \nabla_{u} h(u^{(l)}).
	\end{eqnarray}
To find the maximizer $u^*$ given $\beta, \sigma^2$, we iterate the MM update \eqref{eqn:model1-maximize-mu} until convergence. Note that the indicated matrix inverse in \eqref{eqn:model1-maximize-mu} only needs to be done once and remains constant through the iterations. 

\item  
Updating $\beta$ given $\sigma^2$ and $u^*$ is a regular logistic regression with offset $Z u^*$. We invoke a similar MM update as above
\begin{equation}
\label{eqn:model1-update-beta}
\beta^{(t+1)}  =  \beta^{(t)} + \left( 0.25 X^TX \right)^{-1} X^T (y - p^{*}).
\end{equation}
Again the matrix inverse $\left( 0.25 X^TX\right)^{-1}$ only needs to be done once.

\item To update $\sigma^2$ given $\beta$ and $u^*$, the minorization \eqref{eqn:logdet-minorization} leads to the surrogate function
\begin{eqnarray}
& & g(\sigma^2 \mid \sigma^{2(t)}) = - \frac 12 \sum_{i=1}^m \frac{\|u_i^*\|_2^2}{\sigma_i^2}  \nonumber \\
& & \quad - \frac 12 \sum_{i=1}^m \sigma_i^2\tr \left\{\left(\sum_i \sigma_i^{2(t)} Z_i Z_i^T+W^{*-1}\right)^{-1}Z_iZ_i^T\right\} + c^{(t)},
\end{eqnarray}
where $c^{(t)}$ is a constant irrelevant to optimization. Maximization of $g(\sigma^2 \mid \sigma^{2(t)})$ with respect to $\sigma^2$ yields the explicit MM update
	\begin{equation*}
		\sigma_i^{2(t+1)} = \left[\frac{\|u_i^*\|_2^2}{\tr \left\{Z_i^T(\sum_i \sigma_i^{2(t)} Z_i 	Z_i^T+W^{*-1})^{-1}Z_i\right\}}\right]^{\frac 12}.
	\end{equation*}
When $q \ll n$, the Woodbury formula facilitates the inversion
\begin{eqnarray*}
	& & \left( \sum_i \sigma_i^{(t)2} Z_i Z_i^T+W^{*-1} \right)^{-1} \\
	&=& W^* - W^* Z(\sigma) \{I_q + Z(\sigma)^T W^* Z(\sigma)\}^{-1} Z(\sigma)^T W^*,
\end{eqnarray*}
where $Z(\sigma) = (\sigma_1 Z_1, \ldots, \sigma_m Z_m)$. Since the iterate is derived based on MM principle, it possesses the ascent property 
\begin{equation}\label{eqn:Algo1-ascent-property-sigma}
L_{LA}(\sigma^{(t+1)} \mid \beta, u^*) \geq L_{LA}(\sigma^{(t)} \mid \beta, u^*). 
\end{equation}
Detailed proof is presented in the Appendix.
\end{enumerate}

Like the penalized iteratively reweighted least squares (PIRLS) algorithm described in \cite{BatesMachlerBolkerWalker15lme4}, parameter estimates  are determined for a fixed weights matrix $W^*$ and then the weights are updated to the current estimates and the process is repeated. The resulting algorithm is extremely simple to implement. Algorithm \ref{algo:mmla1} summarizes the MM algorithm for parameter estimation of the logistic linear mixed model \eqref{eqn:lvc-model}. Each iteration involves one-step update of $\beta$ and $\sigma^2$. Several more steps of updating $\beta$ and $\sigma^2$ give similar results in practice. 

	\begin{algorithm}[h!]
		\SetKwInOut{Input}{Input}\SetKwInOut{Output}{Output}
		\Input{$y$, $X$, $Z_1, \ldots, Z_m$}
		\Output{MLE $\hat \beta$, $\hat \sigma_1^2, \ldots, \hat 	\sigma_m^2$}
		Initialize $\beta^{(0)}, \sigma_i^{(0)}>0$, $i=1,\ldots,m$ \;
		\Repeat{objective value converges}{
		$u^{*} \gets \arg \max_{u} \ h(u\mid \sigma^{2(t)}, \beta^{(t)})$ \;
		$p^{*} \gets 1/\left\{1+\exp \left(-X\beta^{(t)} -Zu^*\right)\right\}$ \;
		$\beta^{(t+1)} \gets \beta^{(t)} + \left( 0.25 X^TX\right)^{-1} X^T (y - p^{*})$ \;
$p^{*} \gets 1/\left\{1+\exp \left(-X\beta^{(t+1)} -Zu^*\right)\right\}$ \;
$W^{*} \gets \text{diag}\left\{p^{*}(1-p^{*})\right\}$		\;
		$\sigma_i^{2(t+1)} \gets \left[\frac{\|u_i^*\|_2^2}{\tr \left\{Z_i^T(\sum_i \sigma_i^{2(t)} Z_i 	Z_i^T+W^{*-1})^{-1}Z_i\right\}}\right]^{\frac 12}, \quad i = 1,\ldots, m$ \;
		}
		\caption{MMLA1 - a MM algorithm to maximize the Laplace approximation of likelihood for model \eqref{eqn:lvc-model}.}
		\label{algo:mmla1}
	\end{algorithm}

\noindent {\bf 3.2. Model formulation 2}

In Laplace approximated log-likelihood \eqref{eqn:laplace-likelihood}, we have $\sigma_i$ in the denominator, thus it cannot be combined with  penalized estimation which will shrink some of $\sigma_i$s to zero. 
Therefore we consider another reparameterization of model \eqref{eqn:lvc-model} by assuming that $\eta$ takes the form 
\begin{equation}\label{eqn:lvc-repam}
\eta = X \beta + \sigma_1Z_1 u_1 + \cdots + \sigma_mZ_m u_m,
\end{equation}
where $u_i \sim N(0_{q_i}, I_{q_i})$ are independent. 
Let $u = (u_1^T, \ldots, u_m^T)^T \in \mathcal{R}^q$ 
be the concatenated random effects and $Z = \left(Z_1, \ldots, Z_m\right) \in \mathcal{R}^{n \times q}$, $q = \sum_{i=1}^m q_i$. Then $\eta = X \beta + ZD u$, where $D = \text{blkdiag}\left(\sigma_1I_{q_1}, \dots,\sigma_mI_{q_m} \right)$ and the complete log-likelihood is
\begin{eqnarray*}
h(u \mid \beta, \sigma)
	=\sum_j \left\{y_j \eta_j - \ln (1 + e^{\eta_j})\right\} - \frac 12\|u\|_2^2 + \text{ terms without } u.
\end{eqnarray*}
Given current iterate $(\beta, \sigma)$, let $u^*$ be the maximizer of $h$ and $\eta^* = X \beta + ZD u^*$. Then the approximated log-likelihood is
\begin{eqnarray}
	& & L_{\text{LA}}(\beta, \sigma) \nonumber\\
	&=& h(u^* \mid \beta, \sigma) - \frac 12 \ln \det \nabla^2 \{- h(u^* \mid \beta, \sigma)\} \nonumber\\
	&=& \sum_j \left\{y_j \eta_j^* - \ln \left(1 + e^{\eta_j^*} \right) \right\}  -\frac 12 \|u^*\|_2^2- \frac 12 \ln \det \left( D^TZ^T W^* ZD + I_q \right) \nonumber\\
	&=& \sum_j \left\{y_j \eta_j^* - \ln \left(1 + e^{\eta_j^*} \right) \right\}  -\frac 12 \|u^*\|_2^2  - \frac 12 \ln \det \left( W^{*-1} + \sum_i \sigma_i^2 Z_i Z_i^T \right)  \nonumber\\
	& & - \frac 12 \ln \det W^*+ \text{terms without} \ \beta , \sigma^2. \label{eqn:laplace-likelihood2}
\end{eqnarray}
Detailed derivations of the above approximated log-likelihood can be found in the Appendix.
Maximizing $h(u\mid \beta, \sigma)$ follows similar MM updates as in \eqref{eqn:model1-maximize-mu}. Given $\sigma^2$ and $\beta$, $u^*$ can be found through MM iterates
\begin{eqnarray*}
	u^{(l+1)}  = u^{(l)} +  \left\{ 0.25 (ZD)^TZD+ I_q\right\}^{-1} \nabla_{u} h(u^{(l)}\mid \beta, \sigma^2)
\end{eqnarray*}
until convergence,
where $\nabla_{u} h(u^{(l)}\mid \beta, \sigma^2) =D^TZ^T (y - p) - u^{(l)}$. Updating $\beta$ given $u^*$ and $\sigma^2$ is the same as update in \eqref{eqn:model1-update-beta}. 

Updating $\sigma^2$ given $\beta$ and $u^*$ depends on three minorizations, which differ from the first reparameterization. Quadratic minorization implies that 
\begin{eqnarray} 
- 1^T\ln \left(1 + e^{\eta^*} \right) &\geq &  -p^{(t)T}\left(\eta^* - \eta^{*(t)}\right) - \frac 18 \Vert\eta^* - \eta^{*(t)}\Vert_2^2 + c^{(t)}\nonumber\\
 & =& -p^{(t)T}ZDu^* - \frac 18 \Vert Z(D - D^{(t)})u^*\Vert_2^2 + c^{(t)}, \label{eqn:quadratic-inequal}
\end{eqnarray}
where $c^{(t)}$ is an irrelevant constant, $p^{(t)}$ is a vector with the $j$th element equal to $e^{\eta_j^{*(t)}} / \left(1 + e^{\eta_j^{*(t)}} \right)$ and $\eta_j^{*(t)}$ is the $j$th element of $\eta^{*(t)} = X\beta + ZD^{(t)}u^* $.  The Cauchy inequality implies that 
\begin{eqnarray}
- \Vert Z(D - D^{(t)})u^*\Vert_2^2  &=& - \left\Vert\sum_{i=1}^mZ_iu_i^*(\sigma_i - \sigma_i^{(t)}) \right\Vert_2^2 \nonumber \\
&\geq & - \left\{\sum_{j=1}^n\sum_{i=1}^m (Z_iu_i^*)_{j}^2\right\}\sum_{i=1}^m (\sigma_i - \sigma_i^{(t)})^2, \label{eqn:cauchy-inequal}
\end{eqnarray}
where $(Z_iu_i^*)_{j}$ is the $j$th element of vector $Z_i\mu_i^*$. Combining \eqref{eqn:quadratic-inequal}, \eqref{eqn:cauchy-inequal} and \eqref{eqn:logdet-minorization} gives the overall minorization function
\begin{eqnarray}
	g(\sigma\mid\sigma^{(t)}) &=& \sum_{i=1}^m \sigma_i\left(y - p^{(t)}\right)^TZ_iu_i^* - \frac 18 \left\{\sum_{j=1}^n\sum_{i=1}^m (Z_iu_i^*)_{j}^2\right\}\sum_{i=1}^m (\sigma_i - \sigma_i^{(t)})^2 \nonumber \\
& &  - \frac 12\sum_{i=1}^m \sigma_i^2\tr \left\{ \left(\sum_i \sigma_i^{2(t)} Z_i Z_i^T+W^{*-1} \right)^{-1}Z_iZ_i^T\right\} + c^{(t)},
\end{eqnarray}
where $\sigma_i$ are nicely separated and only involve quadratic terms. Maximization of $g(\sigma\mid\sigma^{(t)})$ results the following update
\begin{equation}\label{eqn:model2-update-sigma}
\sigma^{(t+1)}_i = \frac{\left(y - p^{(t)}\right)^TZ_iu_i^* + \frac 14 \left\{\sum_{j=1}^n\sum_{i=1}^m (Z_iu_i^*)_{j}^2\right\}\sigma_i^{(t)}}{\tr \left\{\left(\sum_i \sigma_i^{2(t)} Z_i Z_i^T+W^{*-1}\right)^{-1}Z_iZ_i^T\right\} + \frac 14 \left\{\sum_{j=1}^n\sum_{i=1}^m (Z_iu_i^*)_{j}^2\right\}}.
\end{equation}
To account for the non-negative constraint of $\sigma$, at each iteration we set $\sigma_i^{(t+1)} = \text{max}\left(0, \sigma_i^{(t+1)}\right)$. Algorithm \ref{algo:mmla2} summarizes the MM algorithm for model formulation 2 defined in \eqref{eqn:lvc-repam}.
\begin{algorithm}[h!]
		\SetKwInOut{Input}{Input}\SetKwInOut{Output}{Output}
		\Input{$y$, $X$, $Z_1, \ldots, Z_m$}
		\Output{MLE $\hat \beta$, $\hat \sigma_1^2, \ldots, \hat 	\sigma_m^2$}
		Initialize $\beta^{(0)}, \sigma_i^{(0)}>0$, $i=1,\ldots,m$ \;
		\Repeat{objective value converges}{
		$D^{(t)} = \text{diag}\left(\sigma^{(t)}_11_{q_1}, \dots,\sigma_m^{(t)}1_{q_m} \right)$ \;
		$u^{*} \gets \arg \max_{u} \ h(u\mid \sigma^{2(t)}, \beta^{(t)})$ \;
		$p^{(t)} \gets 1/\left\{1+\exp \left(-X\beta^{(t)} -ZD^{(t)}u^*\right)\right\}$ \;
		$\beta^{(t+1)} \gets \beta^{(t)} + \left( 0.25 X^TX\right)^{-1} X^T (y - p^{(t)})$ \;
$p^{(t)} \gets 1/\left\{1+\exp \left(-X\beta^{(t+1)} -ZD^{(t)}u^*\right)\right\}$ \;
$W^{*} \gets \text{diag}\left\{p^{(t)}(1-p^{(t)})\right\}$		\;
		$\sigma_i^{2(t+1)} \gets \text{max}\left[0, \frac{\left(y - p^{(t)}\right)^TZ_iu_i^* + \frac 14 \left\{\sum_{j=1}^n\sum_{i=1}^m (Z_iu_i^*)_{j}^2\right\}\sigma_i^{(t)}}{\tr \left\{\left(\sum_i \sigma_i^{2(t)} Z_i Z_i^T+W^{*-1}\right)^{-1}Z_iZ_i^T\right\} + \frac 14 \left\{\sum_{j=1}^n\sum_{i=1}^m (Z_iu_i^*)_{j}^2\right\}}\right], \quad i = 1,\ldots, m$ \;
		}
\caption{MMLA2 - a MM algorithm to maximize the Laplace approximation of likelihood  for model \eqref{eqn:lvc-repam}.}
		\label{algo:mmla2}
\end{algorithm}

\noindent {\bf 3.3. MM algorithm for maximizing the penalized approximated likelihood}

For variance component selection, we consider the penalization approach using lasso penalty. 
Since the minorization function of $\sigma$ derived in second model formulation is a quadratic function of $\sigma$, it meshes well with penalized estimation. 
Other penalties such as the adaptive lasso \citep{Zou2006Adaptive} and smoothly clipped absolute deviation (SCAD) \citep{Fan2001SCAD} lead to similar algorithms.

The lasso penalized approximated log-likelihood is 
\begin{equation} \label{eqn:logl-lasso}
 - L_{\text{LA}}(\beta, \sigma)  + \lambda \sum_{i=1}^m \vert\sigma_i\vert.
\end{equation}
Finding $u^*$ to maximize $h(u\mid \beta, \sigma)$ and updating $\beta$ are the same as described in algorithm \ref{algo:mmla2}. The only difference lies in the update of $\sigma$ given $u^*$ and $\beta$ in \eqref{eqn:model2-update-sigma}, which now becomes
\begin{eqnarray}
\sigma_i^{(t+1)} &=& \underset{\sigma_i}{\text{arg min}} \ \sigma_i^2 \left[\frac 12\tr \left\{(\sum_i \sigma_i^{2(t)} Z_i Z_i^T+W^{*-1})^{-1}Z_iZ_i^T\right\} + \frac 18 \left\{\sum_{j=1}^n\sum_{i=1}^m (Z_iu_i^*)_{j}^2\right\} \right] \nonumber\\ 
&& \qquad \qquad - \sigma_i \left[\left(y - p^{(t)}\right)^TZ_iu_i^* + \frac 14 \left\{\sum_{j=1}^n\sum_{i=1}^m (Z_iu_i^*)_{j}^2\right\} \sigma_i^{(t)}\right] + \lambda \vert\sigma_i\vert  \\
&=& ST\left(z_i, \gamma_i\right), \nonumber
\end{eqnarray}
where 
\begin{equation}
ST(z, \gamma )  =  \underset{x}{\text{arg min}} \frac 12(x-z)^2 +  \gamma\vert x\vert = \text{sng}(z)\left(\vert z\vert -  \gamma\right)_{+}
\end{equation}
is the soft-thresholding operator and 
\begin{eqnarray*}
z_i &=& \frac{\left(y - p^{(t)}\right)^TZ_iu_i^* + \frac 14 \left\{\sum_{j=1}^n\sum_{i=1}^m (Z_iu_i^*)_{j}^2\right\}\sigma_i^{(t)}}{\tr \left\{(\sum_i \sigma_i^{2(t)} Z_i Z_i^T+W^{*-1})^{-1}Z_iZ_i^T \right\} + \frac 14 \left\{\sum_{j=1}^n\sum_{i=1}^m (Z_iu_i^*)_{j}^2\right\}}, \\
\gamma_i &=& \frac{\lambda}{\tr \left\{(\sum_i \sigma_i^{2(t)} Z_i Z_i^T+W^{*-1})^{-1}Z_iZ_i^T \right\} + \frac 14\left\{\sum_{j=1}^n\sum_{i=1}^m (Z_iu_i^*)_{j}^2\right\}}.
\end{eqnarray*}

\noindent {\bf 3.4. Choice of regularization parameter}

The best $\lambda$ can be selected over a grid using Akaike information criterion (AIC), Bayesian information criterion (BIC), or cross-validation. Here we consider AIC and BIC. Since it is hard to evaluate the log likelihood function, we replace it by its Laplace approximation. Specifically, we use
\begin{eqnarray*}
\text{BIC}(\lambda) &=& -2L_{\text{LA}}(\hat{\beta}, \hat{\sigma}^2) + \log(n)\times\text{df}(\lambda) \\
\text{AIC}(\lambda) &=& -2L_{\text{LA}}(\hat{\beta}, \hat{\sigma}^2) + 2\times\text{df}(\lambda), 
\end{eqnarray*}
where $\text{df}(\lambda)$ is the number of non-zeros in $\hat{\sigma}^2(\lambda)$. In the following simulation studies, we compare AIC and BIC on variance component selection.

\setcounter{section}{4} 
\setcounter{equation}{0} 
\noindent {\bf 4. Simulation studies}

\noindent {\bf 4.1. Random effects ANOVA}	
	
In this section we compare the estimation error and runtime of the MM algorithms (MMLA1 and MMLA2) to three different implementations: (1) the {\tt glmer()} function in the popular {\tt lme4} package in {\sc R} \citep{BatesMachlerBolkerWalker15lme4} (2) {\tt glmm()} function in the {\tt glmm} package in {\sc R} \citep{KnudsonGLMM} and (3) {\tt stan\_glmer()} function in the {\tt rstanarm} package in {\sc R} \citep{stan}. {\tt glmer()} fits a generalized linear mixed-effects model and the default (nAGQ=1) uses Laplace approximation to approximate the original log-likelihood. {\tt glmm()} calculates and maximizes the Monte Carlo likelihood approximation (MCLA) \citep{Geyer1990} to find Monte Carlo maximum likelihood estimates (MCMLEs) \citep{SungGeyer07MCMLE} for the fixed effects and variance components. {\tt rstanarm} package is an {\sc R} interface to the Stan C++ library for Bayesian estimation. {\tt stan\_glmer()} adds independent prior distributions on the regression coefficients as well as priors on the covaraince matrices of the group-specific parameters and perform Bayesian inference via MCMC. 

We simulated data from the following two-way ANOVA model with crossed random effects
\begin{eqnarray*}
&&P(y_{ijk} =1) = 1 / (\exp(-\eta_{ijk})) \\
	&&\eta_{ijk} = x_1\beta_1 + x_2\beta_2 + x_3\beta_3 + \alpha_i + \gamma_j + (\alpha \gamma)_{ij}, \\
	 && i = 1, \dots, 5, j = 1,\dots, 5, k = 1, \dots c,
\end{eqnarray*}
where $\alpha_i \sim N(0,\sigma_{\alpha}^2)$, $\gamma_j \sim N(0,\sigma_{\gamma}^2)$ and $(\alpha\gamma)_{ij} \sim N(0, \sigma_{\alpha\gamma}^2)$ are jointly independent. Here $i$ indexes levels in factor 1, $j$ indexes levels in factor 2, and $k$ indexes observations in the $(i,j)$-combination. This corresponds to $m=3$ variance components. Table \ref{table:glmm-la} displays the results when there are $a=b=5$ levels of each factor, the number of observations $c$ in each combination of factor levels varies from 2 to 200, and the true parameter values are $(\beta_1,\beta_2,\beta_3,\sigma_{\alpha}^2,\sigma_{\gamma}^2,\sigma_{\alpha\gamma}^2) = (0.6, 1.0, -1.0, 0.5, 0.9, 0.3)$. For each scenario, we simulated 50 replicates. The sample size was $n= abc$ for each replicate. Therefore the largest model in Table \ref{table:glmm-la} involves covariance matrix of size $5000 \times 5000$. For $c=100$ and $200$, we omit the results of {\tt glmm} and {\tt rstanarm} since they take too much time when sample size gets larger and the whole simulation takes more than a week to complete. 

We made the following observations. Two MM algorithms (MMLA1 and MMLA2) have very close results, but MMLA2 takes longer time to converge than MMLA1, especially when the number of groups $c$ is large. This is what we expected since the surrogate function derived in  MMLA2 involves two more layers of minorizations, which result in slower convergence. The {\tt glmer()} function failed to converge in many replicates when $c=2$ and produced much worse estimates than MM algorithms. For other values of $c$, {\tt glmer()} delivered estimates comparable to MM algorithm but was $3 \sim 4$ fold slower than MMLA1. {\tt glmm()} and {\tt stan\_glmer()} are much slower since they involve sampling and their estimation performance are not good. The core algorithm in {\tt glmer()} is coded in C and extensively utilizes sparse linear algebra. Our MM algorithms are implemented in the high-level Julia language and ignore sparsity structure. Although it is hard to draw conclusions based on implementations in different languages, this example clearly demonstrates the efficiency and scalability of the MM algorithms for GLMM estimation. 

\addtolength{\tabcolsep}{-1.2pt}   
\begin{table}[!h]
	\caption{Comparison of the MM algorithms with two different parameterizations (MMLA1 and MMLA2) and the {\tt glmer()} function (with {\tt nAGQ=1}) in the {\tt lme4} package, {\tt rstanarm} package, and {\tt glmm} package. Standard errors are given in parentheses. Results for {\tt rstanarm} and {\tt glmm} with $c=100, 200$ are not reported because the simulation takes more than 1 week.}
	\centering
	{\scriptsize
	\begin{tabular}{rrrrrrrrr}
	\hline
		    c & Method  & runtime & $\beta_1 (0.6)$ & $\beta_2 (1.0)$ & $\beta_3 (-1.0)$ & $\sigma_{\alpha}^2 (0.5)$ & $\sigma_{\gamma}^2 (0.9)$ & $\sigma_{\alpha\gamma}^2 (0.3)$ \\
\hline
  2 		  & MMLA1 &  0.19(0.55) &  0.68(0.51) &  1.08(0.43) &
               -0.92(0.51) &  0.52(0.91) &  1.03(1.55) &  0.22(0.37)\\
		  & MMLA2 &  0.14(0.12) &  0.68(0.51) &  1.08(0.43) &
               -0.92(0.51) &  0.52(0.91) &  1.04(1.56) &  0.22(0.37)\\
		 & lme4  &  0.46(0.37) &  2.83(7.22) &  3.52(7.39) &
      -2.42(4.04) & 187(753) & 108(580) & 558(2049) \\
      & rstanarm  &  8.15(0.49) &  0.91(0.69) &  1.42(0.45) &
       -1.20(0.58) &  1.38(1.32) &  2.14(2.23) &  2.60(1.86) \\
                 & glmm & 23.95(45.66) &  0.64(0.53) &  0.91(0.55) &
        -0.76(0.59) &  1.54(3.13) &  0.03(0.07) &  0.06(0.14) \\
[0.1cm]
		  8 		  & MMLA1 &  0.10(0.03) &  0.55(0.21) &  0.96(0.24) &
               -0.98(0.20) &  0.36(0.33) &  0.96(0.94) &  0.34(0.34)\\
		  & MMLA2 &  0.17(0.08) &  0.55(0.21) &  0.96(0.24) &
               -0.98(0.20) &  0.36(0.33) &  0.96(0.94) &  0.34(0.34)\\
		 & lme4  &  0.37(0.10) &  0.60(0.23) &  1.04(0.27) &
      -1.07(0.22) &  0.42(0.38) &  1.15(1.13) &  0.47(0.48) \\
      & rstanarm  & 21.85(1.15) &  0.61(0.24) &  1.05(0.27) &
       -1.09(0.22) &  0.68(0.44) &  1.48(1.20) &  0.72(0.53) \\
                 & glmm & 224.53(492.52) &  0.46(0.17) &  0.82(0.24) &
        -0.85(0.17) &  0.78(1.50) &  0.02(0.03) &  0.04(0.08) \\
[0.1cm]
		 50 		  & MMLA1 &  0.19(0.10) &  0.58(0.07) &  1.01(0.08) &
               -1.00(0.08) &  0.52(0.43) &  0.96(0.81) &  0.31(0.16)\\
		  & MMLA2 &  1.65(0.52) &  0.58(0.07) &  1.01(0.08) &
               -1.00(0.08) &  0.52(0.43) &  0.94(0.72) &  0.31(0.16)\\
		 & lme4  &  0.92(0.12) &  0.59(0.07) &  1.03(0.08) &
      -1.02(0.09) &  0.54(0.45) &  1.01(0.86) &  0.32(0.17) \\
       & rstanarm  & 198.38(26.88) &  0.59(0.07) &  1.04(0.08) &
       -1.02(0.09) &  0.82(0.58) &  1.37(0.92) &  0.42(0.21) \\
                 & glmm & 3613.26(2272.85) &  0.48(0.09) &  0.86(0.12) &
        -0.84(0.12) &  0.88(1.39) &  0.04(0.06) &  0.04(0.07) \\
[0.1cm]
		 100 		  & MMLA1 &  0.58(0.18) &  0.61(0.06) &  1.01(0.06) &
               -1.00(0.06) &  0.65(0.46) &  0.94(0.61) &  0.30(0.11)\\
		  & MMLA2 &  4.28(0.78) &  0.61(0.06) &  1.01(0.06) &
               -1.00(0.06) &  0.67(0.44) &  0.91(0.54) &  0.30(0.11)\\
		 & lme4  &  1.49(0.18) &  0.62(0.06) &  1.02(0.06) &
      -1.01(0.06) &  0.67(0.47) &  0.97(0.63) &  0.31(0.12) \\
       & rstanarm &  --- & --- & --- & --- & --- & --- & ---\\
       & glmm  & --- & --- & --- & --- & --- & --- & ---\\
[0.1cm]
		 200 		  & MMLA1 &  0.98(0.16) &  0.60(0.04) &  0.99(0.04) &
               -0.99(0.04) &  0.45(0.33) &  0.92(0.62) &  0.29(0.12)\\
		  & MMLA2 & 13.49(3.42) &  0.60(0.04) &  0.99(0.04) &
               -0.99(0.04) &  0.50(0.33) &  0.91(0.51) &  0.29(0.12)\\
		 & lme4  &  2.76(0.33) &  0.60(0.04) &  1.00(0.04) &
      -1.00(0.04) &  0.46(0.33) &  0.94(0.63) &  0.30(0.13) \\
             & rstanarm &  --- & --- & --- & --- & --- & --- & ---\\
       & glmm  & --- & --- & --- & --- & --- & --- & ---\\	  
		\hline
	\end{tabular}
}
	\label{table:glmm-la}
\end{table}
\addtolength{\tabcolsep}{-1.2pt}

\noindent {\bf 4.2. Genetic example}

In this section, we use a genetic example to demonstrate the performance of variable selection using our algorithm derived in Section 3.3.  
Consider the QTL mapping example introduced in Section 1
\begin{eqnarray*}
g(\mu)  &=&   X \beta + G \gamma,
\end{eqnarray*}
where $G$ is an $n \times k$ genotype matrix for $k$ variants of interest, $g(\mu) = \text{logit}\left(\mu\right)$, $\beta$ are fixed effects, and $\gamma$ are random genetic effects with $\gamma \sim \text{Normal}\left(0, \sigma^2I_k\right)$. The response $y$ is an $n \times 1$ vector of binary trait measurements with mean $\mu$. One way to identify important genes is to test the null hypothesis $\sigma^2=0$ for each region separately and then adjust for multiple testing \citep{Lee14RVSurvey}. Here we consider the joint model for all regions instead of marginal tests
\begin{eqnarray}
	g(\mu) &=& X \beta + s_1^{-1/2} G_1 \gamma_1 + \cdots + s_m^{-1/2}G_m \gamma_m, \label{eqn:genetic-joint}
\end{eqnarray}
where $\gamma_i \sim N(0,\sigma_{i}^2 I)$ and select the variance components $\sigma_i^2$ via the penalization \eqref{eqn:logl-lasso}. Here $s_i$ is the number of variants in region $i$, and the weights $s_i^{-1/2}$ put all variance components on the same scale.

In this simulation study, we use the genetic data from COPDGene exome sequencing study \citep{Regan10COPD}, which has $399$ subjects and genotype information of 16,610 genes. The covariate matrix $X$ contains {\tt intercept}, {\tt age}, {\tt sex}, and the top 3 principal components in the mean effects. We consider four experimental settings for sparse random effects. In all the examples, we set $\beta = \left(0.1, -1.0, 0.8, -0.3, -1.2, 1.5\right)$ and randomly select $m$ genes $G_i$, $i=1,\ldots,m$, from the COPD data. 
\begin{itemize}
\item Setting 1: $\sigma^2 = \left(5.0, 7.5, 10.0, 0^T_{m-3}\right)^T$ with $m$ varying from 5, 10, 20, 100
\item Setting 2: $\sigma^2 = \left(10, 15, 20, 0^T_{m-3}\right)^T$ with $m$ varying from 5, 10, 20, 100
\item Setting 3: $\sigma^2 = \left(5,6,7,8,9,10, 0^T_{m-6}\right)^T$ with $m$ varying from 10, 20, 40, 100
\item Setting 4: $\sigma^2 = \left(10, 12, 14, 16, 18, 20, 0^T_{m-6}\right)^T$ with $m$ varying from 10, 20, 40, 100
\end{itemize}

We use mean squared error (MSE) = $\Vert\hat{\beta}- \beta\Vert^2$ to evaluate the performance of fixed effect estimation. Four measures are used to assess the variable selection performance: the number of truly non-zero variance components that are selected as non-zero variance components (denoted as ``True Positive''), the number of truly zero variance components that are selected as non-zero variance components (denoted as ``False Positive'') , the frequency of exactly selecting the correct variance components (denoted by ``Exact''), and the frequency of over-selecting variance components (denoted by ``Over''). In each experimental setting, 100 data sets are simulated from the model, and we report the average performance over the 100 runs for both AIC and BIC. Table \ref{tab:vs-3-1}, \ref{tab:vs-3-2}, \ref{tab:vs-6-1} and \ref{tab:vs-6-2} summarize the results for the above four settings. We can see that our proposed method for variable selection does  a good job in identifying the significant random effects. For example, under Setting 1 and Setting 2 for different $m$, our method based on both AIC and BIC can identify the truly significant random effects 97\% $\sim$ 99\% of the time with AIC more prone to over-selection than BIC. Setting 3 and Setting 4 are more challenging since they involve a larger number of random effects. But our method can still identify the non-zero random effect 96\% of the time under $m$ = 10 when using AIC.

\begin{table}[!h]
	\caption{Estimation and selection results for Setting 1.}
	\centering
	{\footnotesize
	\begin{tabular}{llrrrrr}\\   
	\hline
    & & & \multicolumn{4}{c}{Variance components selection}  \\
    \cline{4-7}
	m & Criteria & MSE ($\beta$) & True Positive (3)& False Positive (0) & Exact & Over \\
\hline
5

                  & AIC &  0.31(0.20) &  2.98 &  0.33 & 66\% & 32\% \\
                  & BIC &  0.31(0.20) &  2.98 &  0.15 & 84\% & 14\% \\
                  [0.25cm] 
10

                  & AIC &  0.27(0.17) &  2.96 &  1.14 & 26\% & 70\% \\
                  & BIC &  0.29(0.18) &  2.93 &  0.61 & 50\% & 44\% \\
                 [0.25cm]
20

                  & AIC &  0.26(0.16) &  2.96 &  2.01 & 11\% & 86\% \\
                  & BIC &  0.29(0.17) &  2.87 &  1.25 & 17\% & 72\% \\                                          
 				 [0.25cm]
                  
100 
                  & AIC  &  0.30(0.18) &  2.74 &  2.95 &  4\% & 71\% \\
       			  & BIC &  0.38(0.21) &  2.50 &  0.57 & 27\% & 24\% \\                                          
\hline
	\end{tabular}
	}
    \label{tab:vs-3-1} 
\end{table}

 \begin{table}[!h]
 	\caption{Estimation and selection results for Setting 2.}
 	\centering
	{\footnotesize
 	\begin{tabular}{llrrrrr}\\
 	\hline   
     & & & \multicolumn{4}{c}{Variance components selection}  \\
     \cline{4-7}
 	m & Criteria & MSE ($\beta$) & True Positive (3)& False Positive (0) & Exact & Over \\
 \hline	
5

                  & AIC &  0.37(0.22) &  2.99 &  0.40 & 63\% & 36\% \\
                  & BIC &  0.38(0.22) &  2.99 &  0.22 & 79\% & 20\% \\ [0.25cm]
10

                  & AIC &  0.33(0.20) &  2.98 &  1.17 & 28\% & 70\% \\
                  & BIC &  0.36(0.21) &  2.98 &  0.68 & 44\% & 54\% \\ [0.25cm]
20

                  & AIC &  0.34(0.22) &  2.98 &  1.60 & 25\% & 74\% \\
                  & BIC &  0.38(0.24) &  2.95 &  0.85 & 39\% & 58\% \\                                            
[0.25cm]
               
100 
				  & AIC  &  0.37(0.19) &  2.83 &  3.31 &  3\% & 80\% \\
        			  & BIC  &  0.48(0.22) &  2.68 &  0.61 & 38\% & 30\% \\
 \hline
 	\end{tabular}
	}
     \label{tab:vs-3-2} 
 \end{table}
 
 \begin{table}[!h]
	\caption{Estimation and selection results for Setting 3.}
	\centering
	{\footnotesize
	\begin{tabular}{llrrrrr}\\  
	\hline 
    & & & \multicolumn{4}{c}{Variance components selection}  \\
    \cline{4-7}
	m & Criteria & MSE ($\beta$) & True Positive (6)& False Positive (0) & Exact & Over \\
\hline
10

                  & AIC &  0.78(0.30) &  5.96 &  0.84 & 34\% & 62\% \\
                  & BIC &  0.83(0.32) &  5.66 &  0.33 & 54\% & 25\% \\[0.25cm]
20

                  & AIC &  0.73(0.27) &  5.88 &  1.49 & 15\% & 73\% \\
                  & BIC &  0.82(0.32) &  5.56 &  0.48 & 41\% & 32\% \\[0.25cm]    
 40

                  & AIC &  1.04(0.33) &  5.68 &  1.96 & 15\% & 57\% \\
                  & BIC &  1.17(0.37) &  4.96 &  0.74 & 29\% & 27\% \\                                
[0.25cm]
                  
100               
				  & AIC &  0.85(0.34) &  5.40 &  2.54 &  2\% & 48\% \\
                  & BIC &   0.98(0.38) &  4.82 &  0.63 & 12\% & 14\% \\                              
\hline
	\end{tabular}
	}
    \label{tab:vs-6-1} 
\end{table}

\begin{table}[!h]
	\caption{Estimation and selection results for Setting 4.}
	\centering
	{\footnotesize
	\begin{tabular}{llrrrrr}\\   
	\hline
    & & & \multicolumn{4}{c}{Variance components selection}  \\
    \cline{4-7}
	m & Criteria & MSE ($\beta$) & True Positive (6)& False Positive (0) & Exact & Over \\
\hline	
10

                  & AIC &  1.06(0.32) &  5.97 &  0.85 & 32\% & 65\% \\
                  & BIC &  1.09(0.32) &  5.91 &  0.56 & 45\% & 47\% \\[0.25cm]
20

                  & AIC &  1.02(0.34) &  5.96 &  1.36 & 15\% & 81\% \\
                  & BIC &  1.07(0.34) &  5.92 &  0.70 & 38\% & 54\% \\[0.25cm]
40

                  & AIC &  1.44(0.39) &  5.74 &  1.82 & 13\% & 62\% \\
                  & BIC &  1.51(0.40) &  5.54 &  0.85 & 29\% & 39\% \\                 
[0.25cm]
                  
100  
				  & AIC  &  1.18(0.42) &  5.72 &  2.10 &  6\% & 68\% \\
       			  & BIC  &  1.29(0.43) &  5.29 &  0.71 & 21\% & 22\% \\                
\hline
	\end{tabular}
	}
    \label{tab:vs-6-2} 
\end{table}

\setcounter{section}{5} 
\setcounter{equation}{0} 
\noindent {\bf 5. Real data analysis}

\begin{figure}
\centering
	\includegraphics[scale=0.5]{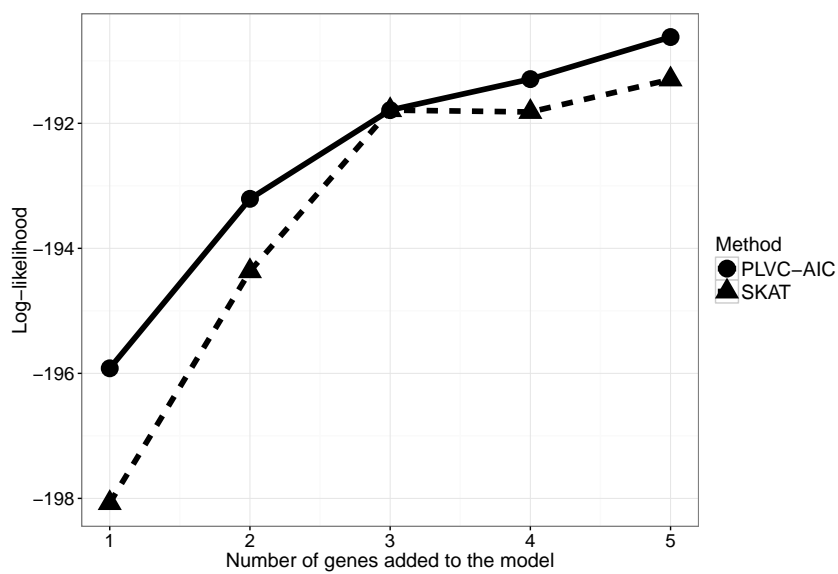}
\caption{Log-likelihood evaluation with top 5 genes selected by PLVC-AIC and SKAT added to the model respectively in an association study of 200 genes and the complex trait {\tt smoke}.}
\label{fig:copd_compare}
\end{figure}

In this real data analysis, we still use the data from COPDGene exome sequencing study described in the above simulated genetic example. The binary trait is smoke or not (denoted as {\tt smoke}). There are $399$ individuals with 646,125 genetic variants in 16,610 genes. The covariates include {\tt age}, {\tt sex}, and the top 3 principal components. Because the number of genes is too large, we first screen the 16,610 genes down to 200 genes according to their marginal p-values from the Sequence Kernel Association Test (SKAT) and then carry out penalized estimation of the 200 variance components in the joint model \eqref{eqn:genetic-joint}. This is similar to the sure independence screening strategy for selecting mean effects \citep{FanLv2008}. AIC selects 16 genes, while BIC criteria selects only one gene ``AFAP1L2''. Table \ref{table:copd-aic-result_threshold_10} lists the top 5 genes selected using AIC criteria (PLVC-AIC) and SKAT. We can see that the top 3 genes selected using both methods are the same but with different order. To compare the selection performance between SKAT and PLVC-AIC, we evaluate the log-likelihood of model \eqref{eqn:genetic-joint} with the top 5 genes listed in Table \ref{table:copd-aic-result_threshold_10} entering the model one by one. To evaluate the log-likelihood, we use the R package {\tt bernor}  which implements the Monte Carlo approximation method described in \cite{SungGeyer07MCMLE}. 
From Figure \ref{fig:copd_compare}, we can see that the log-likelihood with genes selected by PLVC-AIC is above that of SKAT, which in some sense indicates that genes selected by PLVC-AIC explain more variability in the model. 

Besides, we also compare the prediction performance between the top 5 genes selected by PLVC-AIC and SKAT. 
We evaluate the prediction performance using model (\ref{eqn:genetic-joint}) by including the genotype matrix $G_i$ of the corresponding selected genes similar to what is done in \cite{Wu2011RareSKAT}. For example, if the genotype matrix of the top $k$ genes selected are $G_{h_1}, G_{h_2}, \dots, G_{h_k}$, then the predictive model becomes 
\begin{eqnarray*}
	g(\mu) &=& X \beta + s_{h_1}^{-1/2} G_{h_1} \gamma_1 + \cdots + s_{h_k}^{-1/2}G_{h_k} \gamma_{k} = X^*\beta^*, 
\end{eqnarray*}
where $X^* = \left(X, s_{h_1}^{-1/2} G_{h_1}, \dots, s_{h_k}^{-1/2}G_{h_k}\right)$ and $\beta^* = \left(\beta^T, \gamma_1^T,\dots, \gamma_k^T\right)$. This is the ordinary logistic regression model that can be used for prediction. 
Table \ref{table:copd-prediction} summarizes the prediction performance using 5-fold cross validation as the top 5 genes selected by both methods entering the model (\ref{eqn:genetic-joint}) one by one. We can see that on average the model with genes selected by PLVC-AIC performs slightly better than SKAT in terms of prediction. The penalization approach for selecting variance components warrants further theoretical study. This real data analysis demonstrates that the proposed simple MM algorithm scales to high-dimensional problems.

\begin{table}[!h]
\caption{Top 5 genes selected by (1) the lasso penalized variance component model \eqref{eqn:logl-lasso} with AIC criterion (PLVC-AIC) and (2) SKAT in an association study of 200 genes and the binary trait {\tt smoke}.}
\begin{center}
{\footnotesize
\begin{tabular}{c|lcc|lcc}
\hline
& \multicolumn{3}{c|}{PLVC-AIC}  & \multicolumn{3}{c}{SKAT} \\
    \cline{2-7} 
No. & Gene & Marginal p-value & \# Variants & Gene & Marginal p-value & \# Variants\\
\hline
 1&  AFAP1L2 & $6.0 \times 10^{-4}$ & 18 & KIAA1377 & $5.7 \times 10^{-4}$ & 14\\
 2&  RREB1    & $6.0 \times 10^{-4}$  & 18 & RREB1 & $6.0 \times 10^{-4}$ & 18 \\
 3&  KIAA1377 & $5.7 \times 10^{-4}$ & 14 & AFAP1L2 & $6.0 \times 10^{-4}$ & 18 \\
 4&  PSG5    & $3.7 \times 10^{-3}$ & 11 & KARS & $6.1 \times 10^{-4}$ & 15 \\
 5&  TDRD1  & $1.2 \times 10^{-3}$ & 14 & PZP & $1.0 \times 10^{-3}$ & 21 \\
\hline
\end{tabular}
}\label{table:copd-aic-result_threshold_10}
\end{center}
\end{table}

\begin{table}[!h]
\caption{5-fold cross validation performance on prediction accuracy with top 5 genes selected by PLVC-AIC and SKAT added to the model respectively in an association study of 200 genes and the complex trait {\tt smoke}.}
\begin{center}
{\footnotesize
\begin{tabular}{c|c|c}
\hline
& \multicolumn{2}{c}{Prediction accuracy}\\
\cline{2-3}
No. of genes entered into model & PLVC-AIC &  SKAT \\
     
\hline
          1 & 79.4\%(6.2\%) & 78.2\%(4.6\%) \\
		  2 & 79.9\%(6.0\%) & 77.9\%(2.9\%) \\
		  3 & 80.7\%(4.1\%) & 80.7\%(4.1\%) \\
		  4 & 81.7\%(2.3\%) & 80.7\%(5.4\%) \\
		  5 & 81.4\%(3.4\%) & 78.7\%(5.8\%) \\

\hline
\end{tabular}
}\label{table:copd-prediction}
\end{center}

\end{table}

\setcounter{section}{6} 
\setcounter{equation}{0} 
\noindent {\bf 6. Discussion}

This paper discusses two MM algorithms for variance component estimation and selection in the logistic linear mixed model.  The algorithms are simple to implement and scale to models with a large number of variance components. Other extensions are possible. This paper only considers the binary response. The extension of the algorithm MMLA1 to the Poisson count data is straightforward with almost identical derivation. There is work on selecting fixed effects in GLMMs in literature. Here we only focus on random effects selection. Our algorithms can be easily extend to selecting fixed and random effects simultaneously. 
We leave a thorough study of these topics to future research.

\vskip 14pt
\noindent {\large\bf Acknowledgements}

We sincerely thank the AE and two reviewers for their valuable comments and suggestions for improving the manuscript. HZ is partially supported by National Science Foundation (NSF) grant DMS-1310319 and National Institutes of Health (NIH) grants HG006139, GM105785, and GM53275. JZ is supported by NIH grant K01DK106116 and partially by DHS-14-GPD-044-000-98 and U01AI122275. COPDGene study is supported by NIH R01 HL089856 and R01 HL089897.  The whole exome sequencing was supported by the NHLBI Exome Sequencing Project. 
\par
\vskip 14pt
\noindent {\large\bf Appendix}

\noindent {\bf 1. Derivation of approximated log-likelihood in \eqref{eqn:laplace-likelihood}.}
\begin{equation*}
L_{\text{LA}}(\beta, \sigma)
		= h(u^* \mid \beta, \sigma^2) - \frac 12 \ln \det \nabla^2 \left\{- h(u^* \mid \beta, \sigma^{2})\right\},
\end{equation*}
where \begin{equation*}
h(u^* \mid \beta, \sigma^2) =  \sum_j \left\{y_j \eta_j^* - \ln \left(1 + e^{\eta_j^*} \right) \right\} - \frac 12 \sum_{i=1}^m q_i \ln \sigma_i^2 - \frac n2 \ln 2\pi -  \frac 12 \sum_{i=1}^m \frac{\|u_i^*\|_2^2}{\sigma_i^2}.\end{equation*}
The gradient and Hessian of $h(u \mid \beta, \sigma)$ at $u=u^*$ are
	\begin{eqnarray*}
		\nabla_{u} h(u \mid \beta, \sigma^2)_{\mid u=u^*} &=& Z^T (y - p^*) - \begin{pmatrix} \sigma_1^{-2} u^*_1 \\ \vdots \\ \sigma_m^{-2} u^*_m \end{pmatrix}, \\
		\nabla_{u}^2 h(u \mid \beta, \sigma^2)_{\mid u=u^*} &=& - \left\{ Z^T W^*Z + \text{blkdiag}(\sigma_1^{-2}I_{q_1}, \ldots, \sigma_m^{-2} I_{q_m}) \right\}, 
	\end{eqnarray*}
where $p^* = \left(p^*_1, \dots, p^*_n\right)^T$ with $p^*_j = e^{\eta_j^*} / (1 + e^{\eta_j^*})$ and $W^* = \diag(w^*)$ is a diagonal matrix with entries
	\begin{eqnarray*}
		w^*_j = p^*_j (1 - p^*_j) = \frac{e^{\eta_j^*}}{\left( 1 + e^{\eta_j^*} \right)^2}.
	\end{eqnarray*}
Therefore, 
\begin{eqnarray}
		L_{\text{LA}}(\beta, \sigma)
		&=& \sum_j \left\{y_j \eta_j^* - \ln \left(1 + e^{\eta_j^*} \right) \right\} - \frac 12 \sum_{i=1}^m q_i \ln \sigma_i^2 - \frac n2 \ln 2\pi - \frac 12 \sum_{i=1}^m \frac{\|u_i^*\|_2^2}{\sigma_i^2} \nonumber\\
		& & - \frac 12 \ln \det \left\{ Z^T W^* Z + \text{blkdiag}(\sigma_1^{-2}I_{q_1}, \ldots, \sigma_m^{-2} I_{q_m}) \right\} \label{eqn:mmla-derive-1}.
\end{eqnarray}
Using the matrix determinant lemma, we have 
\begin{eqnarray}
 && \ln \det \left\{ Z^T W^* Z + \text{blkdiag}(\sigma_1^{-2}I_{q_1}, \ldots, \sigma_m^{-2} I_{q_m}) \right\} \nonumber\\
 &=& \ln \det \left( W^{*-1} + \sum_i \sigma_i^2 Z_i Z_i^T \right)  + \ln \det  \left( \text{blkdiag}(\sigma_1^{-2}I_{q_1}, \ldots, \sigma_m^{-2} I_{q_m})  \right) + \ln \det W^* \nonumber\\
 &=& \ln \det \left( W^{*-1} + \sum_i \sigma_i^2 Z_i Z_i^T \right) - \sum_{i=1}^m q_i \ln \sigma_i^2 + \ln \det W^*. \label{eqn:mmla-derive-2}
\end{eqnarray}
Substitute \eqref{eqn:mmla-derive-2} to \eqref{eqn:mmla-derive-1} gives
\begin{eqnarray*}
		L_{\text{LA}}(\beta, \sigma)
		&=& \sum_j \left\{y_j \eta_j^* - \ln \left(1 + e^{\eta_j^*} \right) \right\}- \frac 12 \sum_{i=1}^m \frac{\|u_i^*\|_2^2}{\sigma_i^2} \\
		& & - \frac 12 \ln \det \left( W^{*-1} + \sum_i \sigma_i^2 Z_i Z_i^T \right) -\frac 12\ln \det W^* + \text{constant term},
\nonumber	
\end{eqnarray*}
where the constant term equals $- \frac n2 \ln 2\pi$.

\noindent {\bf 2. Derivation of approximated log-likelihood in \eqref{eqn:laplace-likelihood2}.}
\begin{equation*}
 L_{\text{LA}}(\beta, \sigma)
	= h(u^* \mid \beta, \sigma) - \frac 12 \ln \det \nabla^2 \{- h(u^* \mid \beta, \sigma^{2})\},
\end{equation*}
where 
\begin{eqnarray*}
h(u \mid \beta, \sigma^2)
	=\sum_j \left\{y_j \eta_j - \ln (1 + e^{\eta_j})\right\} - \frac 12\|u\|_2^2 - \frac n2 \ln 2\pi.
\end{eqnarray*}
The gradient and Hessian $h(u \mid \beta, \sigma)$ at $u=u^*$ are
	\begin{eqnarray*}
		\nabla_{u} h(u \mid \beta, \sigma)_{\mid u=u^*} &=& D^TZ^T (y - p^*) - u^*, \\
		\nabla_{u}^2 h(u \mid \beta, \sigma)_{\mid u=u^*} &=& - \left( D^TZ^T W^*Z D+ I_{q} \right), 
	\end{eqnarray*}
where $p^* = \left(p^*_1, \dots, p^*_n\right)^T$ with $p^*_j = e^{\eta_j^*} / (1 + e^{\eta_j^*})$ and $W^* = \diag(w^*)$ is a diagonal matrix with entries
	\begin{eqnarray*}
		w^*_j = p^*_j (1 - p^*_j) = \frac{e^{\eta_j^*}}{\left( 1 + e^{\eta_j^*} \right)^2}.
	\end{eqnarray*}
Using the matrix determinant lemma, we have 
\begin{eqnarray}
\ln \det \left( D^TZ^T W^*Z D+ I_{q} \right) 
 =\ln \det \left( W^{*-1} + \sum_i \sigma_i^2 Z_i Z_i^T \right)  + \ln \det W^*. \nonumber
\end{eqnarray}
Therefore,
 \begin{eqnarray}
	& & L_{\text{LA}}(\beta, \sigma) \nonumber\\
	&=& \sum_j \left\{y_j \eta_j^* - \ln \left(1 + e^{\eta_j^*} \right) \right\}  -\frac 12 \|u^*\|_2^2 - \frac n2 \ln 2\pi - \frac 12 \ln \det \left( D^TZ^T W^* ZD + I_q \right) \nonumber\\
	&=& \sum_j \left\{y_j \eta_j^* - \ln \left(1 + e^{\eta_j^*} \right) \right\}  -\frac 12 \|u^*\|_2^2  - \frac 12 \ln \det \left( W^{*-1} + \sum_i \sigma_i^2 Z_i Z_i^T \right) \nonumber\\
	& & - \frac 12 \ln \det W^*+ \text{constant term}, \nonumber
\end{eqnarray} 
where the constant term equals $- \frac n2 \ln 2\pi$.

\noindent {\bf 3. Proof of ascent property in  \eqref{eqn:Algo1-ascent-property-sigma}.}
\begin{proof}
From \eqref{eqn:laplace-likelihood}, the approximated log-likelihood is
\begin{eqnarray*}
L_{\text{LA}}(\beta, \sigma)
&=& \sum_j \left\{y_j \eta_j^* - \ln \left(1 + e^{\eta_j^*} \right) \right\}- \frac 12 \sum_{i=1}^m \frac{\|u_i^*\|_2^2}{\sigma_i^2} \\
		& & \quad  - \frac 12 \ln \det \left( W^{*-1} + \sum_i \sigma_i^2 Z_i Z_i^T \right) - \frac 12 \ln \det W^* + \text{ terms without } \beta , \sigma^2,
\end{eqnarray*}
where $u^*$  is the maximizer of $h(u\mid \beta, \sigma)$, $\eta^* = X\beta + Zu^*$ and $W^* = \diag(w^*)$ is a diagonal matrix with entries
	\begin{eqnarray*}
		w^*_j = p^*_j (1 - p^*_j) = \frac{e^{\eta_j^*}}{\left( 1 + e^{\eta_j^*} \right)^2}.
	\end{eqnarray*}
Thus	
\begin{eqnarray*}
L_{\text{LA}}(\sigma \mid \beta, u^*)
&=& - \frac 12 \sum_{i=1}^m \frac{\|u_i^*\|_2^2}{\sigma_i^2} 
		- \frac 12 \ln \det \left( W^{*-1} + \sum_i \sigma_i^2 Z_i Z_i^T \right) + c,
\end{eqnarray*}
where $c = \sum_j \left\{y_j \eta_j^* - \ln \left(1 + e^{\eta_j^*} \right) \right\} - \frac 12 \ln \det W^* - \frac n2 \ln 2\pi$ is a constant not involving $\sigma$. 

The minorization \eqref{eqn:logdet-minorization} leads to the surrogate function of $L_{\text{LA}}(\sigma \mid \beta, u^*)$
\begin{eqnarray*}
& & g(\sigma^2 \mid \sigma^{2(t)}) = - \frac 12 \sum_{i=1}^m \frac{\|u_i^*\|_2^2}{\sigma_i^2}  
- \frac 12 \sum_{i=1}^m \sigma_i^2\tr \left\{\left(\sum_i \sigma_i^{2(t)} Z_i Z_i^T+W^{*-1}\right)^{-1}Z_iZ_i^T\right\} + c^{(t)},
\end{eqnarray*}
where $c^{(t)}$ is a constant irrelevant to optimization. Since $\sigma^{2(t+1)} = (\sigma_1^{2(t+1)}, \dots, \sigma_m^{2(t+1)})$ with 
	\begin{equation*}
		\sigma_i^{2(t+1)} = \left[\frac{\|u_i^*\|_2^2}{\tr \left\{Z_i^T(\sum_i \sigma_i^{2(t)} Z_i 	Z_i^T+W^{*-1})^{-1}Z_i\right\}}\right]^{\frac 12}
	\end{equation*}
maximizes the surrogate function $g(\sigma^2 \mid \sigma^{2(t)})$, we have the following inequality satisfied
\begin{equation*}
L_{LA}(\sigma^{(t+1)} \mid \beta, u^*) \geq  g(\sigma^{2(t+1)} \mid \sigma^{2(t)} )  \geq g(\sigma^{2(t)} \mid \sigma^{2(t)} ) = L_{LA}(\sigma^{(t)} \mid \beta, u^*). 
\end{equation*}
Therefore, the iterates possess the ascent property.
\end{proof}


\par

\par\vspace{20mm}
\noindent{\large\bf References}

\begin{description}
\bibitem[Ahn et~al., 2012]{AhnZhangLu012MomentRFLMM}
Ahn, M., Zhang, H.~H., and Lu, W. (2012). Moment-based method for random effects selection in linear mixed
  models.
{\it Statistica Sinica}, {\bf 22}, 1539.

\bibitem[Anderson and Aitkin, 1985]{Anderson1985VC}
Anderson, D.~A. and Aitkin, M. (1985).
Variance component models with binary response: interviewer
  variability.
{\it Journal of the Royal Statistical Society. Series B
  (Methodological)}, 203-210.

\bibitem[Bates et~al., 2015]{BatesMachlerBolkerWalker15lme4}
Bates, D., Mächler, M., Bolker, B., and Walker, S. (2015). Fitting linear mixed-effects models using lme4. {\it Journal of Statistical Software}, {\bf 67}, 1-48.

\bibitem[Bondell et~al., 2010]{Bondell2010JointLMM}
Bondell, H.~D., Krishna, A., and Ghosh, S.~K. (2010). Joint variable selection for fixed and random effects in linear
  mixed-effects models. {\it Biometrics}, {\bf 66}, 1069-1077.

\bibitem[Booth and Hobert, 1999]{BoothHobert99GLMM-MCEM}
Booth, J.~G. and Hobert, J.~P. (1999). Maximizing generalized linear mixed model likelihoods with an
  automated monte carlo em algorithm. {\it Journal of the Royal Statistical Society: Series B (Statistical
  Methodology)}, {\bf 61}, 265-285.

\bibitem[Boyd and Vandenberghe, 2004]{BoydVandenberghe04Book}
Boyd, S. and Vandenberghe, L. (2004). {\it Convex optimization}. Cambridge university press.

\bibitem[Breslow and Clayton, 1993]{BreslowClayton93GLMM}
Breslow, N.~E. and Clayton, D.~G. (1993). Approximate inference in generalized linear mixed models. {\it Journal of the American statistical Association}, {\bf 88}, 9-25.

\bibitem[Cai and Dunson, 2006]{CaiDunson2006BayesianCovSelection}
Cai, B. and Dunson, D.~B. (2006). Bayesian covariance selection in generalized linear mixed models. {\it Biometrics}, {\bf 62}, 446-457.

\bibitem[Che and Xu, 2012]{Che2012QTLGLMM}
Che, X. and Xu, S. (2012).
Generalized linear mixed models for mapping multiple quantitative
  trait loci.
{\it Heredity}, {\bf 109}, 41-49.

\bibitem[Davidian and Gallant, 1992]{Davidian1992GaussianQuadrature}
Davidian, M. and Gallant, A.~R. (1992).
Smooth nonparametric maximum likelihood estimation for population
  pharmacokinetics, with application to quinidine.
{\it Journal of Pharmacokinetics and Pharmacodynamics},
  {\bf 20}, 529-556.

\bibitem[Fan and Li, 2001]{Fan2001SCAD}
Fan, J. and Li, R. (2001). Variable selection via nonconcave penalized likelihood and its oracle
  properties. {\it Journal of the American statistical Association},
  {\bf 96}, 1348-1360.

\bibitem[Fan and Lv, 2008]{FanLv2008}
Fan, J. and Lv, J. (2008). Sure independence screening for ultrahigh dimensional feature space. {\it Journal of the Royal Statistical Society: Series B (Statistical
  Methodology)}, {\bf 70}, 849-911.

\bibitem[Geyer, 1990]{Geyer1990}
Geyer, C. (1990). Likelihood and Exponential Families. PhD thesis, University of Washington.

\bibitem[Groll and Tutz, 2014]{Groll2014VSGLMM}
Groll, A. and Tutz, G. (2014). Variable selection for generalized linear mixed models by l1-penalized estimation. {\it Statistics and computing}, {\bf 24}, 137-154.

\bibitem[Hunter and Lange, 2004]{Hunter2004MMtutorial}
Hunter, D.~R. and Lange, K. (2004).
A tutorial on mm algorithms.
{\it The American Statistician}, {\bf 58}, 30-37.

\bibitem[Ibrahim et~al., 2011]{Ibrahim2011SelectMixedEffect}
Ibrahim, J.~G., Zhu, H., Garcia, R.~I., and Guo, R. (2011). Fixed and random effects selection in mixed effects models. {\it Biometrics}, {\bf 67}, 495-503.

\bibitem[Knudson, 2016]{KnudsonGLMM}
Knudson, C. (2016).
glmm: Generalized Linear Mixed Models via Monte Carlo Likelihood
  Approximation. R package version 1.1.1.

\bibitem[Lange et~al., 2000]{Lange00OptTrans}
Lange, K., Hunter, D.~R., and Yang, I. (2000). Optimization transfer using surrogate objective functions. {\it J. Comput. Graph. Statist.}, {\bf 9}, 1-59. With discussion, and a rejoinder by Hunter and Lange.

\bibitem[Lee et~al., 2014]{Lee14RVSurvey}
Lee, S., Abecasis, G.~R., Boehnke, M., and Lin, X. (2014). Rare-variant association analysis: study designs and statistical
  tests. {\it The American Journal of Human Genetics}, {\bf 95}, 5-23.

\bibitem[McCulloch and Neuhaus, 2001]{Mcculloch2001GLMM}
McCulloch, C.~E. and Neuhaus, J.~M. (2001).
{\it Generalized linear mixed models}. Wiley Online Library.

\bibitem[Pan and Huang, 2014]{Pan2014RandomGLMM}
Pan, J. and Huang, C. (2014). Random effects selection in generalized linear mixed models via
  shrinkage penalty function. {\it Statistics and Computing}, {\bf 24}, 725-738.

\bibitem[Pinheiro and Bates, 1995]{PinheiroBates1995ApproximationsNLM}
Pinheiro, J.~C. and Bates, D.~M. (1995). Approximations to the log-likelihood function in the nonlinear
  mixed-effects model. {\it Journal of computational and Graphical Statistics}, {\bf 4}, 12-35.

\bibitem[Quen{\'e} and Van~den Bergh, 2008]{Quene2008ANOVA}
Quen{\'e}, H. and Van~den Bergh, H. (2008).
Examples of mixed-effects modeling with crossed random effects and
  with binomial data.
{\it Journal of Memory and Language}, {\bf 59}, 413-425.

\bibitem[Regan et~al., 2011]{Regan10COPD}
Regan, E.~A., Hokanson, J.~E., Murphy, J.~R., Make, B., Lynch, D.~A., Beaty,
  T.~H., Curran-Everett, D., Silverman, E.~K., and Crapo, J.~D. (2011). Genetic epidemiology of copd (copdgene) study design. {\it COPD: Journal of Chronic Obstructive Pulmonary Disease},
  {\bf 7}, 32-43.

\bibitem[Rue et~al., 2009]{RueMartinoChopin09INLA}
Rue, H., Martino, S., and Chopin, N. (2009). Approximate bayesian inference for latent gaussian models by using
  integrated nested laplace approximations. {\it Journal of the royal statistical society: Series b (statistical
  methodology)}, {\bf 71}, 319-392.

\bibitem[Schelldorfer et~al., 2014]{Schelldorfer2014GLMMlasso}
Schelldorfer, J., Meier, L., and B{\"u}hlmann, P. (2014). Glmmlasso: an algorithm for high-dimensional generalized linear mixed
  models using L1-penalization. {\it Journal of Computational and Graphical Statistics},
  {\bf 23}, 460-477.

\bibitem[Shun and McCullagh, 1995]{ShunMcCullagh95LA}
Shun, Z. and McCullagh, P. (1995). Laplace approximation of high dimensional integrals. {\it Journal of the Royal Statistical Society. Series B
  (Methodological)}, pages 749-760.

\bibitem[{Stan Development Team}, 2016]{stan}
{Stan Development Team} (2016).
rstanarm: {Bayesian} applied regression modeling via {Stan}.
R package version 2.13.1.

\bibitem[Stroup, 2012]{Stroup2012GLMM}
Stroup, W.~W. (2012).
{\it Generalized linear mixed models: modern concepts, methods and
  applications}. CRC press.
\bibitem[Sung and Geyer, 2007]{SungGeyer07MCMLE}
Sung, Y.~J. and Geyer, C.~J. (2007). Monte carlo likelihood inference for missing data models. {\it The Annals of Statistics}, 990-1011.

\bibitem[Wolfinger, 1993]{Wolfinger93LA}
Wolfinger, R. (1993). Laplace's approximation for nonlinear mixed models. {\it Biometrika}, 791-795.

\bibitem[Wu et~al., 2011]{Wu2011RareSKAT}
Wu, M.~C., Lee, S., Cai, T., Li, Y., Boehnke, M., and Lin, X. (2011).
Rare-variant association testing for sequencing data with the
  sequence kernel association test.
{\it The American Journal of Human Genetics}, {\bf 89}, 82-93.

\bibitem[Yi and Xu, 1999]{Yi1999QTLmapping}
Yi, N. and Xu, S. (1999).
Mapping quantitative trait loci for complex binary traits in outbred
  populations.
 {\it Heredity}, {\bf 82}, 668-676.

\bibitem[Zeger and Karim, 1991]{Zeger1991GLMM}
Zeger, S.~L. and Karim, M.~R. (1991).
Generalized linear models with random effects; a gibbs sampling
  approach.
{\it Journal of the American statistical association},
  {\bf 86}, 79-86.

\bibitem[Zou, 2006]{Zou2006Adaptive}
Zou, H. (2006). The adaptive lasso and its oracle properties. {\it Journal of the American statistical association},
  {\bf 101}, 1418-1429.
\end{description}


\vskip .65cm
\noindent
{\small
Department of Statistics, North Carolina State University,  Raleigh, North Carolina 27695, U.S.A 
\vskip 2pt
\noindent
E-mail: lhu@ncsu.edu
\vskip 2pt

\noindent
Department of Statistics, North Carolina State University,  Raleigh, North Carolina 27695, U.S.A 
\vskip 2pt
\noindent
E-mail: lu@stat.ncsu.edu
\vskip 2pt

\noindent
Department of Epidemiology and Biostatistics, University of Arizona, 
Tucson, AZ 85721-0066, U.S.A
\vskip 2pt
\noindent
E-mail: jzhou@email.arizona.edu
\vskip 2pt

\noindent
Department of Biostatistics, University of California, 
Los Angeles, California 90095-1772, U.S.A
\vskip 2pt
\noindent
E-mail: huazhou@ucla.edu
}
\end{document}